\documentclass[10pt]{article}
\usepackage{amsfonts}
\usepackage{mathrsfs}
\usepackage{amsmath}	
\usepackage{cite}
\usepackage{graphicx}
\usepackage{rotating}
\usepackage{hyperref}
\usepackage{verbatim}
\usepackage[all]{xy}

\csname @addtoreset\endcsname{equation}{section}
\newcommand{\beq}{\begin{equation}}
\newcommand{\eeq}{\end{equation}}
\newcommand{\bea}{\begin{eqnarray}}
\newcommand{\eea}{\end{eqnarray}}
\newcommand{\ba}{\begin{array}}
\newcommand{\ea}{\end{array}}
\newcommand{\bit}{\begin{itemize}}
\newcommand{\eit}{\end{itemize}}
\newcommand{\nn}{\nonumber}

\newcommand{\mezzo}{\frac{1}{2}}
\newcommand{\complesso}{{\ \hbox{{\rm I}\kern-.6em\hbox{\bf C}}}}
\newcommand{\reale}{{\hbox{{\rm I}\kern-.2em\hbox{\rm R}}}}
\newcommand{\uno}{ \,  \raisebox{+0.14em}{{\hbox{{\rm \scriptsize ]}} \raisebox{-0.2em}{\kern-.8em\hbox{1}}}} \, }  %  operatore identit\`a

\newcommand{\p}{\partial}
\renewcommand{\a}{\alpha}
\renewcommand{\b}{\beta}
\newcommand{\g}{\gamma}

\renewcommand{\d}{\delta}

\newcommand{\Er}{{\mathcal{E}}}

\renewcommand{\k}{\kappa}
\renewcommand{\l}{\lambda}

\newcommand{\m}{\mu}

\newcommand{\n}{\nu}
\renewcommand{\r}{\rho}
\newcommand{\s}{\sigma}

\renewcommand{\t}{\theta}

\newcommand{\om}{\omega}
\newcommand{\Om}{\Omega}

\begin{document}

%\begin{comment}

\begin{titlepage}
\begin{flushright}
%hep-th/??????\\
CECS-PHY-14/04
\end{flushright}
\vspace{2.5cm}
\begin{center}
\renewcommand{\thefootnote}{\fnsymbol{footnote}}
{\huge \bf Stationary axisymmetric  spacetimes}
\vskip 5mm
{\huge \bf with a conformally coupled scalar field}
\vskip 30mm
{\large {Marco Astorino\footnote{marco.astorino@gmail.com}}}\\
\renewcommand{\thefootnote}{\arabic{footnote}}
\setcounter{footnote}{0}
\vskip 10mm
{\small \textit{
Centro de Estudios Cient\'{\i}ficos (CECs), Valdivia,\\ 
Chile\\}
}
\end{center}
\vspace{5.5 cm}
\begin{center}
{\bf Abstract}
\end{center}
{Solution generating techniques for general relativity with a conformally (and minimally) coupled scalar field are pushed forward to build a wide class of asymptotically flat, axisymmetric and stationary spacetimes continuously connected to Kerr black hole. This family contains, amongst other things, rotating extensions of the Bekenstein black hole and also its angular and mass multipolar generalisations. Further addition of NUT charge is also discussed.}
\end{titlepage}

%\end{comment}

%\newpage

                        %%%%%%%%%%%%%%%%%%%%%%%%%%%%%%%%%%%%%%%%%%%%%%%%%%%%%%%%%%%%%%%%%%%%%%%
                        %                                                                     %
                        %                             START                                   %
                        %                                                                     %
                        %%%%%%%%%%%%%%%%%%%%%%%%%%%%%%%%%%%%%%%%%%%%%%%%%%%%%%%%%%%%%%%%%%%%%%%

%\tableofcontents

\section{Introduction}
\label{intro}

Fundamental scalar fields have been studied for a long time in gravity and high energy theoretical physics with various aims ranging from   cosmology to the standard model (of particles), scalar-tensor theories and strings. But lately they are enjoying renewed attention after the experimental confirmation of the Higgs scalar field at CERN. Historically the interest in the scalar matter field coupled to general relativity in a conformal invariant way (such as standard Maxwell electromagnetism, in four dimension) have arisen in seventies when Bekenstein had shown that coupling could admit a black hole solution \cite{bekenstein1}, \cite{bekenstein2}. At that time it constituted the first counterexample to the black hole no-hair theorem, which states that all degrees of freedom, in the gravitational collapse forming a black hole, vanish apart from the mass and the angular momentum (and electric charge, in case we are considering also electromagnetic coupling). This black hole,  found by Bocharova, Bronnikov, Melnikov \cite{BBM} and Bekenstein \cite{bekenstein1}, \cite{bekenstein2} (henceforward BBMB), present some issues summarised in \cite{marcoa2}. The main ones are the fact that the spacetime is not stable under linear perturbations \cite{Bronnikov} and the fact that the scalar field is divergent on the event horizon\footnote{In \cite{bekenstein2} is clarified, as suggested by de Witt, that this divergency does not cause any pathological behaviour on physical observables, for example while crossing the horizon there is not potential barrier and tidal forces remain finite.}. Note that in the presence of a cosmological constant the scalar field infinities are hidden behind the event horizon \cite{martinez}, therefore the solution becomes more regular.   \\  
Nevertheless lately there has been some interest in the solution generating techniques for general relativity with a conformally coupled scalar field \cite{hairy}, \cite{charm-conf} and  in its main application, i.e. the rotating generalisation of the BBMB black hole, which is still missing. Some stationary generalisations of BBMB spacetime were produced including acceleration \cite{andres-hideki}; or an external magnetic field  \cite{hairy}, \cite{marcoa2};  or NUT charge \cite{hideki-nut}, \cite{charm-conf}. The main inconvenience shared by these constructions is that they are not asymptotically flat nor have a proper limit to the Kerr black hole. Recently the possibility of having a slowly rotating generalisation of the BBMB metric has been discussed in  \cite{hideki-nut} and \cite{joel}. \\
The aim of this paper is to fill this gap, i.e. to exploit and enhance the techniques developed in \cite{hairy} to find a general asymptotically flat, axisymmetric and  stationary rotating family of metrics for the conformally coupled scalar matter, which include as a static limit the BBMB black hole.  This is done in section \ref{sec-bbmb}.  For this purpose we have to integrate the methods of \cite{hairy}, based on the Ernst formalism \cite{ernst1}, with the HKX transformation \cite{kinnersley-6} originally developed to  add rotation to static axisymmetric spacetimes in general relativity, while preserving the asymptotic flatness. For example, these are the best transformations for generating the Kerr black hole form the Schwarzschild one. Basically we want to generalise some of the results presented in \cite{dietz-h} and \cite{quevedo} in the presence of a conformally coupled scalar field, multipolar metrics are considered in section \ref{multi-section}. \\
As explained in \cite{hairy}, when a scalar field is conformally coupled with general relativity\footnote{We are not considering the cosmological constant here because a solution generating technique in that case is not available at the moment \cite{marcoa-lambda}. } the most generic axisymmetric and stationary spacetime is not modelled by the Lewis-Weyl-Papapetrou metric.  Therefore, in order to take advantage of the Weyl coordinates and of  the integrability of the system, we shift from the conformally coupled theory ($\textrm{CC}$) to the minimally coupled one ($\textrm{MC}$), thanks to a conformal transformation of the metric.  Then we  make use of the explicit symmetries of the minimally coupled theory, which allow us to perform an $\hat{HKX}$ transformation that is able to generate rotation, and finally we come back to the conformally coupled theory, thanks to a conformal transformation (inverse with respect to the first one). With this procedure we can generate a $HKX$ transformation also in the conformally coupled theory.  Pictorially this is illustrated in the following figure:
\begin{displaymath}
\xymatrix{
MC  \ar[d]_{\hat{HKX}} &
CC \ar[l]_{\Om^{-1}} \ar[d]^{HKX=\Om \circ \hat{HKX} \circ \Om^{-1}} \\
MC \ar[r]_\Om
&  CC}
\end{displaymath}

To be more precise, let us  consider the action for general relativity with a conformally coupled scalar field $\Psi$ :
\beq  \label{action}
                       I[g_{\m\n}, \Psi] =    \int d^4x  \sqrt{-g} \left[ \frac{\textrm{R}}{16 \pi G} - \frac{1}{2} \p_\m \Psi \p^\m \Psi - \frac{\textrm{R}}{12} \Psi^2 \right]  \ \ \ .
\eeq
Extremising the action with respect to the metric $g_{\m\n}$ yields the Einstein field equations, while extremising with respect to the scalar field $\Psi$ gives the scalar field equation:
\bea  \label{field-eq}
                       &&  \textrm{R}_{\m\n} - \mezzo  \textrm{R}  g_{\m\n}  = 8 \pi G \  \left[  \p_\m \Psi \p_\n \Psi - \mezzo g_{\m\n} \p_\s \Psi \p^\s \Psi + \frac{1}{6}\left(g_{\m\n} \Box -\nabla_\m \nabla_\n + G_{\m\n} \right) \Psi^2\right] ,       \\
               %        &&  T_{\m\n} = \p_\m \Psi \p_\n \Psi - \mezzo g_{\m\n} \p_\s \Psi \p^\s \Psi + \frac{1}{6}\left(g_{\m\n} \Box -\nabla_\m \nabla_\n + G_{\m\n} \right) \Psi^2\ \ ;\nn \\
                       &&  \Box \Psi - \frac{1}{6} \textrm{R} \Psi  = 0  \ \ \quad .
\eea
We now focus on a subclass of stationary axisymmetric spacetimes, that contain the  BBMB black hole in the static case, which can be generally written as 
\beq \label{lwp-conf}
       ds^2= \Omega \left\{  -f \left( dt - \om d\varphi \right)^2 + f^{-1} \left[ \rho^2 d\varphi^2 + e^{2\gamma}  \left( d\rho^2 + dz^2 \right) \right] \right\}  \quad ,
\eeq
where all the functions $f,\gamma,\omega$ and $\Omega$ depend on the $(\r,z)$ coordinates only. $\Omega$ is the conformal factor that relates the minimally coupled theory to the conformally coupled one (\ref{action}):
\beq    \label{conf-tr}
             \Omega(\r,z):= \left[1-\frac{4 \pi G}{3}\Psi^2(\r,z) \right]^{-1}  \quad .
\eeq
Actually any solution of general relativity with a minimally coupled scalar field ($\hat{g} , \hat{\Psi}$), whose action is
\beq  \label{minimal-action}
                       \hat{I}[\hat{g}_{\m\n}, \hat{\Psi}] =  \int d^4x  \sqrt{-\hat{g}}\ \left[  \frac{\hat{R}}{16 \pi G}   -  \ \mezzo \nabla_\m \hat{\Psi} \nabla^\m \hat{\Psi} \right]  \  
\eeq
and whose field equations are 
\bea  \label{min-field-eq}
                        &&  \hat{ \textrm{R}}_{\m\n} -   \frac{\hat{\textrm{R}}}{2}  \hat{g}_{\m\n} = 8\pi G \left( \p_\m \hat{\Psi} \p_\n \hat{\Psi} - \mezzo \hat{g}_{\m\n} \p_\s \hat{\Psi} \p^\s \hat{\Psi} \right)  \quad ,       \\
         \label{min-field-eq-psi}               &&   \Box \hat{\Psi}=0 \quad ,
\eea
can be mapped into a solution ($g,\Psi$) of the conformally coupled theory (\ref{action}) by the following conformal transformation
\bea 
    \label{conf-tras-psi} \hat{\Psi} &  \longrightarrow & \Psi=\sqrt{\frac{6}{8\pi G}} \tanh \left(\sqrt{\frac{8\pi G}{6}} \hat{\Psi} \right) \ \ , \\ 
    \label{conf-tras-g} \hat{g}_{\m\n} &\longrightarrow & g_{\m \n} = \Omega \ \hat{g}_{\m\n}  \ \ .
\eea
At this point it is possible to use the solution generating technique developed in \cite{hairy} for the theory  (\ref{action}). It consists of building Ernst potentials for the  minimally coupled theory (\ref{minimal-action}) and then uplifting it to the conformally coupled theory by the conformal transformation (\ref{conf-tras-psi})-(\ref{conf-tras-g}).  For generating  purposes usually the best coordinates are the prolate spherical ones ($x,y$), which are related to ($\r,z$) by the following transformations
\beq 
                 \r:= \k  \sqrt{(x^2-1)(1-y^2)} \qquad , \quad  \qquad \qquad z:=\k x y \quad , \quad
\eeq
where $\k$ is a constant. In \cite{hairy} we have learnt that the symmetries of axisymmetric and stationary solutions of standard general relativity are inherited by the conformally coupled theory, so we can also use the improvements of the Ernst technique \cite{ernst1} developed by Hoenselaers, Kinnersley and Xanthopoulos (HKX) in \cite{kinnersley-6} (see also \cite{quevedo}), to generate a stationary version of the  BBMB metric from the static one.  \\
As a starting point we consider the Fisher, Janis, Robinson, Winnicour metric (FJRW), which is a static solution for the minimally coupled theory; in prolate spherical coordinates it can be written as
\beq
            \hat{ds}^2 = - \left(\frac{x-1}{x+1} \right)^\d dt^2 + \left(\frac{x+1}{x-1} \right)^\d \k^2 \left[ dx^2 + \frac{x^2-1}{1-y^2} dy^2 + (x^2-1)(1-y^2) d\varphi^2 \right]   \quad .
\eeq
It is supported by the following scalar field:
\beq \label{scalar-mini}
            \hat{\Psi}_0(x) = \sqrt{\frac{1-\d^2}{16\pi G}} \log\left( x-1 \over x+1 \right) \quad .
\eeq
From this seed metric we can extract its Ernst potential, (since the metric is static and electromagnetically uncharged $\Er=f$)
\beq \label{Er-zip}
         \Er_0 = \left(\frac{x-1}{x+1} \right)^\d  \quad ,
\eeq   
where the distortion (or Zipoy-Voorhees) parameter $\d \in \mathbb{R}$. We recall that for $\d=1$ we have the Schwarzschild spacetime (note that, in this case, the scalar field vanishes), while for $\d=1/2$ we have the BBMB black hole, up to the conformal transformation (\ref{conf-tr}), as explicitly shown in the next section. Note that the scalar field (\ref{scalar-mini}) is not the most general solution of Eq. (\ref{min-field-eq-psi}) but rather just the one giving the BBMB metric; this is our motivation for picking it.  While other possible generalisations of the scalar field (\ref{scalar-mini}) are considered in appendix \ref{app-app-expan-g}. Note also that the HKX transformations do not affect the scalar field, like all transformations inherited by the vacuum symmetries.  \\

\section{Adding rotation to the BBMB Black Hole}
\label{sec-bbmb}

In this section we want to find a stationary generalisation of the BBMB black hole. Thus we apply two rank-zero HKX transformations to the static (therefore real) seed Ernst potential (\ref{Er-zip}), as done in \cite{dietz-h} and \cite{quevedo} for general relativity. The presentation of the HKX  formalism is rather involved and beyond of the scope of the present paper; for a detailed introduction on HKX transformations and their applications to vacuum Weyl metrics see \cite{hoense} and \cite{dietz}. However we can present the action of $N$ rank zero HKX transformations on a static seed Ernst potential $\Er_0$ to get a new stationary potential $\Er$
\beq
                    \Er_0 \longrightarrow \Er = \Er_0 \ \frac{D_-}{D_+}
\eeq
where
\beq \label{N-rank-zero-HKX}
            D_\pm = \textrm{det} \left\{ \d_{ij} +i \frac{\a_k U_k}{2 S(U_k)} \exp \big[ 2 B(U_k) \big]  \left[ \frac{U_ j+ U_k-4U_j U_k z}{U_j S(U_k)  + U_k S(U_j)}  \pm 1  \right]     \right\} \ .
\eeq
This transformation adds $2N$ parameters $\a_k$ and $U_k$ because $j,k=1,2,...,N$. The function $B(U_k)$ satisfies the differential equation \footnote{The differential operator $\overrightarrow{\nabla}$ refers to the flat cylindrical gradient in $(\r,z,\varphi)$ coordinates.} 
\beq
            S(U_k) \overrightarrow{\nabla} B(U_k) (1 - 2 U_k z)  \overrightarrow{\nabla} (\mezzo \log \Er_0) + 2 U_k \rho  \overrightarrow{e}_\varphi \times \overrightarrow{\nabla} (\mezzo \log \Er_0)
 \eeq
with 
\beq
             S^2(U_k) = (1-2 U_k z)^2 + (2 U_k \rho)^2
\eeq
For the two rank-zero HKX transformations $k \in \{1,2\}$, so they add four new constants  $\a_1,\a_2, U_1$ and $U_2$, two of which can be reabsorbed in a coordinate transformation
\beq
          U_1 = - U_2 = \frac{1}{2\k} = U \qquad ,
\eeq
then
$$   S(\pm U) = x \mp y  \quad . $$
By inserting this latter in (\ref{N-rank-zero-HKX}), redefining the constants $\a_1:=\a$ and  $\a_2:=\b$, we get a new rotating (therefore complex) Ernst potential for the stationary version of the FJRW metric:
\beq \label{ernst-potential}
            \Er = \frac{d_-}{d_+} = \frac{\xi-1}{\xi+1}      \qquad \qquad  \text{with} \qquad \xi:=\frac{d_++d_-}{d_+-d_-}  \quad ,
\eeq
where 
\bea
               d_\pm (x,y) &:=&  (x\pm 1)^{\d-1} \left[ x (1-\l \m) + i y (\l + \m) \pm (1+\l\m) \mp i (\l-\m) \right]   \label{dpm}  \quad ,\\
               \lambda(x,y) &:=&  \a (x^2-1)^{1-\d} (x+y)^{2\d - 2}   \label{lambda} \quad ,\\
               \mu(x,y) &:=&  \b (x^2-1)^{1-\d} (x-y)^{2\d - 2}   \label{mu} \quad .
\eea
The two rank-zero HKX transformations add two new independent parameters $\a$ and $\b$, usually called rotation and reflection parameters. In general, for $\d\neq1$, the presence of $\a$ and $\b$, with $\a\neq\b$, may break the equatorial symmetry with respect to the plane $y=0$, while the axisymmetry is always granted by construction through (\ref{lwp-conf}). 
The HKX-transformed  potential generally may have NUT charge, which can spoil the asymptotic flatness of the seed metric. Therefore we perform an additional Ehlers transformation to add another NUT charge, parametrised by $\tau$, which can elide the possible pre-existing one. The Ehlers transformation in terms of $\xi$ consists just in adding a multiplying phase: $\xi \longrightarrow \bar{\xi}= \xi e^{i\tau}$, therefore the final Ernst potential $\bar{\Er}$ reads
\beq \label{ersnt-pot-mono}
         \bar{\Er} = \frac{\bar{\xi}-1}{\bar{\xi}+1} = \frac{(d_++d_-) e^{i\tau} - (d_+-d_-) }{(d_++d_-) e^{i\tau}+(d_+-d_-) }  \quad .
\eeq
The Ernst potential (\ref{ersnt-pot-mono}) represents the stationary rotating version of the FJRW metric, describing a mass monopole, which additionally is asymptotically  flat, or at most NUT. Mass multipolar solutions can also be constructed with the help of the solution generating techniques, this will be done in section \ref{multi-section}. We remember that a spacetime can have both mass multipoles and angular momentum multipoles, but generally these latter vanish in the Newtonian limit.\\
Moreover note that the $\d$ parameter remains a real number in the stationary case as well, and it is not limited to integers as it happens for the standard Tomimatsu-Sato family.

%%%%%%%%%%%%%%%%%%%%%%%%%%%%%%%%%%%%%%%%%%%%%
\subsection{$\a\neq 0$ and $\b=0$} 
\label{b=0}
%%%%%%%%%%%%%%%%%%%%%%%%%%%%%%%%%%%%%%%%%%%%%

For the sake of simplicity let's restrict to the case $\b=0$ in (\ref{mu}), because this is the simplest case containing the Kerr metric. In section  \ref{secB} and appendix \ref{secC} some more general cases are considered.\\
First of all, we want to check that the case $\d=1$ contains the Kerr Black hole. For $\d=1$ the Ernst potential becomes
\beq  \label{ersnt-pot-1}
              \Er_{(1)} = \frac{ ( x + i y \a ) ( \cos \tau + i \sin \tau ) - (1- i \a) }{( x + i y \a ) ( \cos \tau + i \sin \tau ) + (1- i \a) }   \quad .
\eeq
%This is the Ernst potential for the Kerr-NUT spacetime, so 
Then we can cancel the NUT charge by demanding asymptotic flatness. In practice this means we have to impose the following constraints on the parameters
\beq
          \cos \tau = \frac{\k}{m}  \qquad , \qquad  \sin \tau = - \frac{a}{m} \qquad , \qquad  \a = \frac{a}{\k} \qquad , \qquad \k^2=m^2-a^2 \quad . 
\eeq
Hence the Ernst potential for the pure Kerr metric is found:
\beq \label{kerr-pot}
          \Er_{(1)} =   \frac{ \displaystyle x  \frac{\k}{m} + i y \frac{a}{m} - 1}{ \displaystyle x  \frac{\k}{m} + i y \frac{a}{m} + 1}    \quad .
\eeq
In this case the parameters $a$ and $m$ represent, respectively,  the mass and the angular momentum of the Kerr black hole. Note that $\d=1$ implies the vanishing of the scalar field and, as a consequence, the trivialisation of the conformal factor (\ref{conf-tr}), which becomes $\Omega=1$. It means that the Ernst potential (\ref{ersnt-pot-1}), if it is properly cleaned from NUT charges, describes the Kerr metric in both the Einstein and Jordan frames. \\
Since we want to build a stationary version of the BBMB black hole we have to consider $\d=1/2$. In fact, for this value of $\d$, the static BBMB black hole can be obtained by a conformal transformation (\ref{conf-tr}) of the  FJRW spacetime.
So for $\d=1/2$ the Ernst potential becomes
\beq
 \Er_{\left(\frac{1}{2}\right)}  = \frac{\sqrt{x+1} \sin \frac{\tau }{2}  \left[-\text{$\alpha$} (x-1) (y-1)+i \sqrt{x^2-1} (x+y)\right]+\sqrt{x-1} \cos \frac{\tau }{2} \left[\sqrt{x^2-1} (x+y)+i \text{$\alpha $} (x+1) (y+1)\right]}{\sqrt{x-1} \sin \frac{\tau }{2} \left[-\text{$\alpha $} (x+1) (y+1)+i \sqrt{x^2-1} (x+y)\right]+\sqrt{x+1} \cos\frac{\tau }{2} \left[\sqrt{x^2-1} (x+y)+i \text{$\alpha $} (x-1) (y-1)\right]}  \label{ernst12}
\eeq
From the definition of the Ernst potential 
\beq
               \Er := f + i \ h 
\eeq
we can directly infer the $f$ field of the metric (\ref{lwp-conf}), as the real part of (\ref{ernst12}), while $\omega$ can be obtained from the definition of $h$:
\beq \label{nablah}
      \overrightarrow{\nabla} h := - \frac{f^2}{\r} \overrightarrow{e}_\varphi \times \overrightarrow{\nabla} \omega    \qquad .
\eeq
The differential operators in spheroidal coordinates can be taken as follows\footnote{The orthonormal frame is defined by the ordered triad $(\overrightarrow{e}_x, \overrightarrow{e}_y ,\overrightarrow{e}_\varphi)$}
\beq
          \overrightarrow{\nabla} f(x,y) \propto  \frac{\overrightarrow{e}_x}{\k} \sqrt{\frac{x^2-1}{x^2-y^2}} \ \p_x \ f(x,y)  + \frac{\overrightarrow{e}_y}{\k} \sqrt{\frac{1-y^2}{x^2-y^2}} \ \p_y \ f(x,y) \quad ,
\eeq
while the two dimensional line element in spheroidal coordinates is
\beq
         d\rho^2+dz^2=\k^2(x^2-y^2)\left[ \frac{dx^2}{x^2-1} + \frac{dy^2}{1-y^2}  \right] \quad .
\eeq
Up to this point the effects of the minimally coupled scalar field have not been taken into account, because at the level of the Ernst formalism the minimally coupled scalar field is actually decoupled from the Ernst potentials. But to find $\g$ the contributions of the scalar stress energy-tensor are relevant. Usually to obtain $\g$ a quadrature is sufficient, once the other fields are known. In this case, from the $EE^\rho_{\ \rho}$ and $EE^\rho_{\ z}$ components of the Einstein equations ($EE$) in the minimally coupled theory (\ref{min-field-eq}), we have respectively:
\bea
            \p_\rho \g &=& -\frac{1}{4} \frac{f^2}{\rho} \left[ (\p_\r \om)^2-(\p_z \om)^2 \right]  + \frac{1}{4} \frac{\rho}{f^2} \left[ (\p_\r f)^2-(\p_z f)^2 \right] +4\pi G \rho \left[ (\p_\r \hat{\Psi})^2-(\p_z \hat{\Psi})^2 \right]  \label{gr} \\
            \p_z \g    &=&  \frac{\r}{2f^2} (\p_z f)(\p_\r f) - \frac{f^2}{2\r} (\p_z \om)(\p_\r \om) + 8\pi G \rho \ (\p_z \hat{\Psi})(\p_\r\hat{\Psi})  \label{gz} 
\eea
Note that by defining $\g=\g_0+\g_\Psi$, where $\g_0$ is solution for general relativity (when $\Psi=0$), the previous system of partial differential equations (\ref{gr})-(\ref{gz}), thanks to its linearity, reduces to 
\bea
 \p_\rho \g_\Psi &=&   4\pi G \rho \left[ (\p_\r \hat{\Psi})^2-(\p_z \hat{\Psi})^2 \right]  \label{gr-psi} \ \ ,\\
            \p_z \g_\Psi   &=&  8\pi G \rho \ (\p_z \hat{\Psi})(\p_\r\hat{\Psi})  \label{gz-psi} \ \ .
\eea
This means that from any axisymmetric and stationary solution of general relativity  we can generate a new solution for the same theory with the addition of a minimally (or conformally whether properly conformally transformed according to (\ref{conf-tras-psi})-(\ref{conf-tras-g})) coupled scalar field. This can be done just by adding the $\g_\Psi$ contribution given by an harmonic scalar field satisfying (\ref{gr-psi})-(\ref{gz-psi}). The harmonicity is required by the scalar field equation (\ref{min-field-eq-psi}). \\ 
The most general solution of (\ref{min-field-eq-psi}) achievable by separation of variables can be expressed, in prolate spherical coordinates, as an expansion in terms of the Legendre polynomials of the first and second kind (more details in appendix \ref{app-legendre}), denoted $P_n(x)$ and $Q_n(x)$ respectively 
\beq \label{scalar-general}
          \hat{\Psi} = \sum_{n=0}^\infty \left[ a_n Q_n(x) + b_n P_n(x) \right] \left[c_n Q_n(y) + d_n P_n(y)  \right] \quad .
\eeq
Requiring some regularity properties to the scalar field it is possible to constrain the coefficients $a_n, b_n, c_n, d_n$, for instance asking regularity along the symmetry axis ($y=\pm1$) fixes the $c_n=0$ coefficients. In appendix \ref{app-app-expan-g} the first orders of the scalar field  expansion (\ref{scalar-general}) and their contributions to $\g$ are considered, for some suitable boundary conditions.\\
The particular scalar field  (\ref{scalar-mini}) we are focusing on in this paper, i.e. the one that gives the BBMB black hole, can be obtained from the general solution (\ref{scalar-general}) by keeping only the  $a_0$ and $d_0$ coefficients not null, such that  $a_0 d_0 =\sqrt{(1-\d^2)/(16\pi G)}$. In this case is easy to evaluate the scalar field contribution $\g_\Psi$ to the total $\g$; integrating (\ref{gr-psi})-(\ref{gz-psi}) we have
\beq \label{gipsi}
           \g_\Psi = \k_2 - \mezzo (\d^2-1) \log \left( \frac{x^2-1}{x^2-y^2} \right) \quad ,
\eeq
where $\k_2$ is an integrating constant, which can be fixed to fulfil the desired boundary conditions or guarantee the regularity of the metric, such as elementary  asymptotic flatness. To sum up, the resulting fields for the conformally coupled theory and $\d=1/2$ are:
\bea
          f(x,y)&=&\frac{\sqrt{x^2-1}\left[ (x+y)^2 - \a^2 (1-y^2)   \right]}{\cos \tau \left[(x+y)^2-\text{$\alpha $}^2 \left(1+2 x y+y^2 \right)\right]+\text{$\alpha $}^2 \left(x y^2+x+2 y\right)+ (x+2 \text{$\alpha $} \sin \tau )  (x+y)^2} \ \ \   \\
           \omega(x,y)&=&      \kappa\  \frac{  \sin \tau \left[y (x+y)^2+\text{$\alpha $}^2 \left(1-y^2\right) (2 x+y)\right]-2 \text{$\alpha $} y \cos \tau (x+y)^2+2 \text{$\alpha $}^3 \left(1-y^2\right) }{(x+y)^2+\text{$\alpha $}^2 \left(y^2-1\right)} \\
           \g(x,y)&=&   \frac{1}{2} \log \left[\frac{x^2-1}{x^2-y^2}-\frac{\text{$\alpha $}^2 \left(x^2-1\right) \left(1-y^2\right)}{(x+y)^2 \left(x^2-y^2\right)}\right] \label{g_b=0}\\
         \Psi(x) &=&   \sqrt{\frac{3}{ 4 \pi G}} \tanh \left[\frac{1}{4} \log \left(\frac{x-1}{x+1}\right)\right] \label{scalar-conf}
\eea
$\g$ is independent on the NUT parameter $\tau$, but not $\om$. When $\a=0$ we recover the NUT-BBMB metric recently found in \cite{hideki-nut} and \cite{charm-conf}.  In order the metric to be free from the NUT charge we have to ask that $\omega(x,y)\rightarrow0$ at spatial infinity, that is for large $x$. Therefore we have properly fixed the arbitrary integration constant of $\omega$ and furthermore we have  to constrain the $\tau$ parameter as follows
\beq \label{tau-b=0}
          \tau =  \text{ArcTan} \left(- \frac{\a}{\d}  \right)  \quad .
\eeq
Under these flat boundary conditions the functions $f$ and $\om$ simplify into
\bea
           \om &=&\frac{2 \text{$\alpha $}^3 \text{$\kappa $} \left(1-y^2\right) \left(\sqrt{1+4 \text{$\alpha $}^2}+2 x+y\right)}{\sqrt{1+4 \text{$\alpha $}^2} \left[(x+y)^2-\text{$\alpha $}^2 \left(1-y^2\right)\right]} \label{om_b=0}  \\
            f&=& \frac{\sqrt{1+4 \text{$\alpha $}^2} \sqrt{x^2-1} \left[(x+y)^2-\text{$\alpha $}^2 \left(1-y^2\right)\right]}{\sqrt{1+4 \text{$\alpha $}^2} \left[\text{$\alpha $}^2 \left(x y^2+x+2 y\right)+x (x+y)^2\right]+(1+4 \text{$\alpha $}^2) (x+y)^2 -\text{$\alpha $}^2 \left(2 x y+y^2+1\right)} \label{f_b=0}
\eea
The metric is free from conical singularities on the axes of symmetry, since $\lim_{y \rightarrow \pm  1}\g = 0 $ and asymptotically it approaches the Minkowski spacetime. When the parameter $\a=0$ one recovers the BBMB static black hole
\bea \label{bbmb-bh}
           ds^2 \Big|_{\alpha=0} &=&     - \left(  1 - \frac{m}{R}  \right)^2 d\tau^2 + \frac{d R^2}{\left(  1 - \frac{m}{R}  \right)^2} + R^2 \big( d\theta^2 + \sin^2 \theta d\phi^2 \big)  \qquad , \\
           \Psi(R) &=& \pm \sqrt{\frac{3}{4\pi G}} \ \left(1-\frac{R}{m}\right)^{-1} \qquad ,
\eea
where the following relation between the coordinate $x$ and the radial coordinate\footnote{In order to minimise the confusion between the radial coordinate $R$ and scalar curvature invariants, such as the Ricci scalar R,  a different font is used for these latter.} $R$ are used:
\beq \label{xR}
 x:=\frac{R^2}{2m(R-m)} -1 \quad .
\eeq
The double degenerate horizon is located at $R=m$. Therefore, given that $R(x)= m \left(x + 1 \mp \sqrt{x^2-1}\right)$, in terms of the $x$ coordinate, the horizon can be approached, by taking the minus branch, in the limit $x\rightarrow\infty$, while the radial coordinate $R(x)$ points towards spatial infinity  for $x\rightarrow\infty$  when taking the plus branch. \\ 
In the stationary case we do not have a unique criterion to define a radial coordinate as it can be done in the static case requiring, for instance, a spherically  symmetric base manifold. Therefore several possibilities for the radial coordinate can be considered in the rotating case, which physically may not be equivalent everywhere  because of the non differentiability  of the change of coordinates. The fact that the two charts are not  diffeomorphic everywhere stems from the only constraint we have to accomplish: the radial coordinate has to converge to the static one (\ref{xR}) in the non rotating limit ($\a=0$).  The easiest radial coordinate in the rotating case we can define is\footnote{Note that in Kerr case this difficulty is not present because the rotating metric we want to recover is already a known solution, therefore the change of coordinate can be easily established. For instance an alternative radial coordinate, which recovers eq.  (\ref{xR}) in the static limit, can be chosen as   $ x(R):= \frac{R^2}{\k(R-m)} - \frac{2m}{\k}   $; but other choices are possible.}
\beq \label{coor-transf}
         x:= \frac{2R^2}{\k(2R-\k)} - 1   \qquad  \underset{\a\rightarrow 0}{\longrightarrow}   \qquad  \frac{R^2}{2m(R-m)} -1 \quad .
\eeq
The mass and angular momentum can be read from the asymptotic behaviour of the metric, because the scalar field does not contribute to the charges. This is because the scalar field depends only on the radial coordinate and it quickly decays to zero at spatial infinity, and in the Hamiltonian formalism one can see that it is not contributing. 
For large values of the radial coordinate $R$ the metric approaches spatial infinity as
\beq
                  ds^2 \sim  - \left( 1 - \frac{2m}{R} \right)  dt^2 + \left( 1 + \frac{2m}{R} \right) dR^2 + \frac{8 \k^2 \a^3 \sin^2\t}{R \sqrt{1+4\a^2}}  dtd\varphi + R^2( d\t^2 + \sin^2\t \ d\varphi^2 ) +O\left(\frac{1}{R^2}\right) \qquad
\eeq

\begin{comment}
\bea
          g_{tt} &\sim & -\left( 1 - \frac{\k\sqrt{1+4\a^2}}{R} \right) +O\left(\frac{1}{R^2}\right)\\
          g_{t\varphi} &\sim & \sin^2 \t \ \frac{4\a^3\k^2}{R\sqrt{1+4\a^2}} +O\left(\frac{1}{R^2}\right)
\eea
\end{comment}
We now try to adapt the definition of the constant parameters $\k$ and $\a$, as in the Kerr case, while also taking into account the extra constant $\d$:
\beq \label{para}
           \k := \frac{m}{\sqrt{\d^2+\a^2}} \qquad , \qquad \a :=\frac{a \d}{\k}\quad .
\eeq
This value we have chosen for $\k$ coincides, setting $\b=0$, with the more general one given in \cite{yamazaki-81} 
\beq \label{yama-k}
     \k  =   \frac{m(1-\a\b)}{\sqrt{\left[\d(\a\b-1) -2\a\b \right]^2 +(\a-\b)^2}} \quad .
\eeq
With these definitions the mass $M$ and angular momentum $J$ become respectively:
\beq 
           M=m \qquad ,\qquad J= -\frac{8 \alpha^3 m^2}{\left(1+4 \alpha^2 \right)^{3/2}}\quad .
\eeq
%This particular scalar field depending only on the radial coordinate and vanishing at spatial infinity is not contributing neither to the mass or to the angular momentum of the metric, as in the static case.\\
With the help of appendix \ref{app-mome} we can compute the mass and angular multipole moments up to the octupole for the scalar generalisation of the FJRW metric (with $\d=1/2$) defined by equations (\ref{f_b=0}),(\ref{om_b=0}) and (\ref{g_b=0}), in the Einstein frame
\bea
          M_0 &=&  m \qquad \qquad \qquad \ \qquad \qquad \qquad  \ \  \quad  J_0 =  \ 0  \nn \\
          M_1 &=&  -\frac{4 \text{$\alpha $}^2 m^2}{\left(1+4 \text{$\alpha $}^2\right)^{3/2}}  \qquad \ \qquad \qquad \qquad  J_1 = \  -\frac{8 \text{$\alpha $}^3 m^2}{\left(1+4 \text{$\alpha $}^2\right)^{3/2}} \ \ , \nn \\
          M_2 &=& \frac{\left(1 + 8 \text{$\alpha $}^2 -16 \text{$\alpha $}^4\right) m^3}{\left(1+4 \text{$\alpha $}^2\right)^2}  \ \quad \qquad \qquad  J_2 =  \ \frac{16 \text{$\alpha $}^3 m^3}{\left(1+4 \a^2\right)^2} \ \ , \nn  \\ 
          M_3 &=& -\frac{4 \a^2 \left(4 \a^2+3\right) m^4}{\left(1+4 \a^2\right)^{5/2}}   \qquad \qquad \qquad  J_3 = \ \frac{8 \a^3 \left(1-4 \a^2\right) m^4}{\left(1 + 4 \a^2 \right)^{5/2}} \label{mult-ex-b=0} \ \  . 
\eea
A spacetime symmetric with respect to the equatorial plane $y=0$ has a multipolar expansion characterised by even (power of 2) mass poles (monopole, quadrupole, ... ) and odd angular poles (dipole, octupole, ...), such as, for instance, the Kerr spacetime (see appendix \ref{app-mome}). The fact that both even and odd multipole moments are present means that the metric is asymmetric with respect to the equatorial plane.  In fact  odd powers of $y$ are present in the metric functions (\ref{g_b=0})-(\ref{f_b=0}). \\
Moreover the spacetime  (\ref{g_b=0})-(\ref{f_b=0}) presents divergences of  the scalar curvature invariants, such as the Riemann squared $\textrm{R}_{\m\n\s\l} \textrm{R}^{\m\n\s\l} $, which are not covered by a horizon.\\

%%%%%%%%%%%%%%%%%%%%%%%%%%%%%%%%%%%%%
\subsection{$\a = \b\neq 0$}
\label{secB}
%%%%%%%%%%%%%%%%%%%%%%%%%%%%%%%%%%%%%%%%

Interestingly enough the Kerr space-time can be obtained, from the general potential (\ref{ersnt-pot-mono}), in ways other than the one performed in section \ref{b=0}. We will see that, although for $\d=1$ the two constructions coincide, whenever $\d \neq 1$ they give rise to inequivalent Ernst potentials. Therefore we can have different stationary solutions, with the same $\d$, which have the same static limit to the BBMB black hole. This occurs even without adding mass multipoles, which produce extra degeneracy; we will further consider these multipolar generalisation of the FJRW in section \ref{multi-section}.\\
In this section let us consider also a non-null $\m(x,y)$, but for simplicity we set $\b=\a$ in (\ref{mu}), thus we will again keep only one rotation/reflection independent parameter. With these settings fixing $\d=1$ in (\ref{ersnt-pot-mono}) gives us the usual Ernst potential for the Kerr-NUT spacetime\cite{treves}:
\beq \label{ern-pot-a=b}
        \Er = \frac{\xi \ e^{i\tau}-1}{\xi \ e^{i\tau}+1} \ \ \ \qquad \text{with} \qquad \xi=px+iqy \quad .
\eeq
where 
\beq
      p = \frac{1-\a^2}{1+\a^2} \qquad , \qquad  q = \frac{2\a}{1+\a^2}    \quad .
\eeq
Note that $p^2+q^2=1$, as it is expected to be for the Kerr solution. In order to neutralise the NUT charge in this case it is not necessary an Ehlers transformation, we can achieve the same result by simply imposing $\tau=0$. In this way we remain with the Ernst potential for the Kerr black hole, as in (\ref{kerr-pot}) and the $\Er$ simplifies to $d_-/d_+$.\\
Now we will play the same game we have done in the previous section (where $\b=0$), for the FJRW metric with $\d=1/2$, but  under the assumption $\a=\b \neq 0$. In the same way we can derive $\om$ through (\ref{nablah})  and then analyse its asymptotic behaviour for large $x$:
\beq
\om \approx \frac{-4 \text{$\alpha $}^3 \kappa +\text{$\alpha $}^2 \text{$\omega_0$}+\left(3 \text{$\alpha $}^2+1\right) \kappa  y \sin (\tau )-\text{$\omega_0$}}{\text{$\alpha $}^2-1} - \frac{8 \left(\text{$\alpha $}^3 \kappa  \left(y^2-1\right) \cos (\tau )\right)}{\left(\text{$\alpha $}^2-1\right)^2 x} + O\left(1\over x^2\right)
 \eeq
In order to have a good falloff behaviour we require that $\om\longrightarrow0$ at spatial infinity, so we impose 
$$  \om_o=\frac{4 \text{$\alpha $}^3 \kappa }{\text{$\alpha $}^2-1}  \qquad  \qquad \text{and} \qquad \qquad \tau=0 \qquad .$$  
Therefore, as in the $\d=1$ case, when $\a=\b$ the vanishing of the NUT charge is achieved for $\tau=0$. A general expression for $\tau$ in the  case $\a\neq0 \neq\b$ is given in \cite{yamazaki-81} 
\beq \label{tau-yama}
                \tau = \frac{\a-\b}{\d(\a\b-1)-\a\b}
\eeq
Thus, when $\a=\b$, $\tau$ is independent from $\d$, in contrast with what happened in subsection \ref{b=0} . Hence, for these values of the parameters, the asymptotically flat Ernst potential $\Er$ is just $d_-/d_+$.
\beq
\Er = \frac{\sqrt{x^2-1} \left[\text{$\alpha $}^2 (x+1)^2+y^2-x^2\right]-2 i \text{$\alpha $} x^2 y+2 i \text{$\alpha $} y}{(x+1) \left\{\text{$\alpha $}^2-2 i y \text{$\alpha $} \sqrt{x^2-1} +x \left[\text{$\alpha $}^2 (x-2)-x\right]+y^2\right\}} \quad .
\eeq
In order to avoid conical singularity on the axis of symmetry, when integrating $\g$ one has to set the arbitrary integration constant to fulfil 
\beq \label{lit-g}
 \lim_{y\rightarrow \pm 1} \g = 0 \ \ .
\eeq
Finally, after having imposed the elementary flat boundary conditions, we have
\bea
              f &=&\frac{\sqrt{x^2-1} \left[\text{$\alpha $}^4 \left(x^2-1\right)^2+\left(x^2-y^2\right)^2-2 \text{$\alpha $}^2 \left[x^4+x^2 \left(1-3 y^2\right)+y^2\right]\right]}{-2 \text{$\alpha $}^2 \left(x^2-1\right) \left[(x-1) x^2-(3 x+1) y^2\right]+(x+1) \left(x^2-y^2\right)^2+\text{$\alpha $}^4 (x-1)^4 (x+1)} \label{f-rot-bbmb-B} \ \ , \\
      \om   &=& \frac{4 \text{$\alpha $}^3 \kappa  \left(y^2-1\right) \left[2 x^3+x^2-\text{$\alpha $}^2 (x-1)^2 (2 x+1)+y^2\right]}{\left(\text{$\alpha $}^2-1\right) \left[\text{$\alpha $}^4 \left(x^2-1\right)^2+\left(x^2-y^2\right)^2-2 \text{$\alpha $}^2 \left[x^4+x^2 \left(1-3 y^2\right)+y^2\right]\right]} \quad , \\
           e^{2\g} &=& \frac{1}{(\a^2-1)^2} \left[ \frac{x^2-1}{x^2-y^2} -\frac{2 \text{$\alpha $}^2 \left(x^2-1\right) \left(x^4-3 x^2 y^2+x^2+y^2\right)}{\left(x^2-y^2\right)^3} +  \frac{\text{$\alpha $}^4 \left(x^2-1\right)^3}{\left(x^2-y^2\right)^3} \right]  \quad . \label{g-rot-bbmb-B}
\eea
Note that for $\d=1$ both the metrics built here and in the previous section coincide with the Kerr spacetime. But for $\d=1/2$ (and possibly $\forall \ \d\neq1$) the two constructions give rise to inequivalent Ernst potentials. Since the coordinate system $(x,y)$ used for both constructions is the same, i.e. prolate spherical, the two spacetimes are different, as scalar curvature invariants show. Another difference between the two rotating BBMB spacetimes presented in sections \ref{b=0} and \ref{secB} lies in the multipolar expansion. In fact, with the help of appendix \ref{app-mome} and (\ref{yama-k}), we can compute the mass and angular multipole moments up to the octupole, for the metric (\ref{f-rot-bbmb-B})-(\ref{g-rot-bbmb-B}) in the Einstein frame:
\bea
          M_0 &=&  m \qquad \qquad  \qquad \ \qquad \qquad \qquad \ \ \  \qquad \qquad   J_0 =  \ 0  \ \ ,\nn \\
          M_1 &=&  0   \hspace{6.2cm}  J_1 = \   -\frac{16 \text{$\alpha $}^3 m^2}{\left(1+3 \text{$\alpha $}^2\right)^2} \ \ , \nn \\
          M_2 &=& \frac{\left(-5 \text{$\alpha $}^6-69 \text{$\alpha $}^4+9 \text{$\alpha $}^2+1\right) m^3}{\left(1+3 \text{$\alpha $}^2\right)^3} \ \quad \qquad \qquad  J_2 = \  0 \ \ , \nn  \\ 
          M_3 &=& 0   \hspace{6.2cm}  J_3 = \  \frac{8 \k^4 \a^3 \left(\a^2+1\right)}{\left(\a^2-1\right)^4} \label{mul-ex-a=b} \ \  . 
\eea
This multipolar moment expansion differs both qualitatively and quantitatively with respect to the (\ref{mult-ex-b=0}) one. Since the multipole moments considered here are not coordinates dependent, it means that for $\d=1/2$ the metrics constructed in section \ref{b=0} and \ref{secB} are not diffeomorphic, so they describe different spacetimes, in contrast with the case $\d=1$. In particular the multipole expansion  (\ref{mul-ex-a=b})  is typical of metrics that are symmetric with respect to the equatorial plane $y=0$, as can be directly checked in  (\ref{f-rot-bbmb-B})-({\ref{g-rot-bbmb-B}).  \\
For large values of the radial coordinate $R$, as defined in (\ref{coor-transf}) and  taking into account the relation (\ref{yama-k}), the metric approaches spatial infinity as
%\beq
  %                ds^2 \sim  - \left( 1 - \frac{\k}{R} \frac{1+3\a^2}{1-\a^2} \right)  dt^2 + \left( 1 +  \frac{\k}{R} \frac{1+3\a^2}{1-\a^2}\right) dR^2 + \frac{16 \k^2 \a^3 \sin^2\t}{R (1-\a^2)^2}  dtd\varphi + R^2( d\t^2 + \sin^2\t \ d\varphi^2 ) +O\left(\frac{1}{R^2}\right) \qquad \nn
%\eeq
%and taking $\k$ as in (\ref{yama-k})
\beq
                  ds^2 \sim  - \left( 1 - \frac{2m}{R} \right)  dt^2 + \left( 1 + \frac{2m}{R} \right) dR^2 + \frac{64 m^2 \a^3 \sin^2\t}{R (1-\a^2)^2}  dtd\varphi + R^2( d\t^2 + \sin^2\t \ d\varphi^2 ) +O\left(\frac{1}{R^2}\right) \quad . \nn
\eeq
As explained in \cite{carter}, for this class of stationary and axisymmetric spacetimes, the null horizons can be found from the relation
\beq
                   g_{tt} g_{\varphi\varphi} =g_{\varphi t}^2   \qquad \Longrightarrow \qquad \frac{\rho^2}{ (1-\frac{8\pi G}{6} \Psi^2)^2}=0 \quad .
\eeq 
So using the radial coordinate (\ref{coor-transf}), the $R_H=\k/2 $ hypersurface is null, $g^{RR}(R_H)=0$ and it coincides, in the no rotation limit, with the BBMB event horizon
\beq
                    R_H = m \ \frac{1-\a^2}{1+3\a^2}   \quad  \underset{\a\rightarrow 0}{\longrightarrow} \quad m \qquad .
\eeq
Actually the hypersurface $R_H=\k/2$ is double degenerate, as it occurs in the static case, where the geometry is extremal even though the mass parameter is free (and not either the addition of electromagnetic charges to the BBMB black hole can alter its extremality). For a reasonable range of the mass and rotation parameters and radial coordinate $R$,  the scalar curvature invariants, such as the Riemann squared $\textrm{R}_{\m\n\s\l} \textrm{R}^{\m\n\s\l} $, diverge only  at $R=0$, as can be seen in figure \ref{figu1}.  The symmetry axis is located at $y=\pm1$, as it can be checked by the fact that $g_{t\varphi}$ and $g_{\varphi\varphi}$ vanish there. \\
 \begin{figure}[h!]
  \centering
 \includegraphics[scale=0.5]{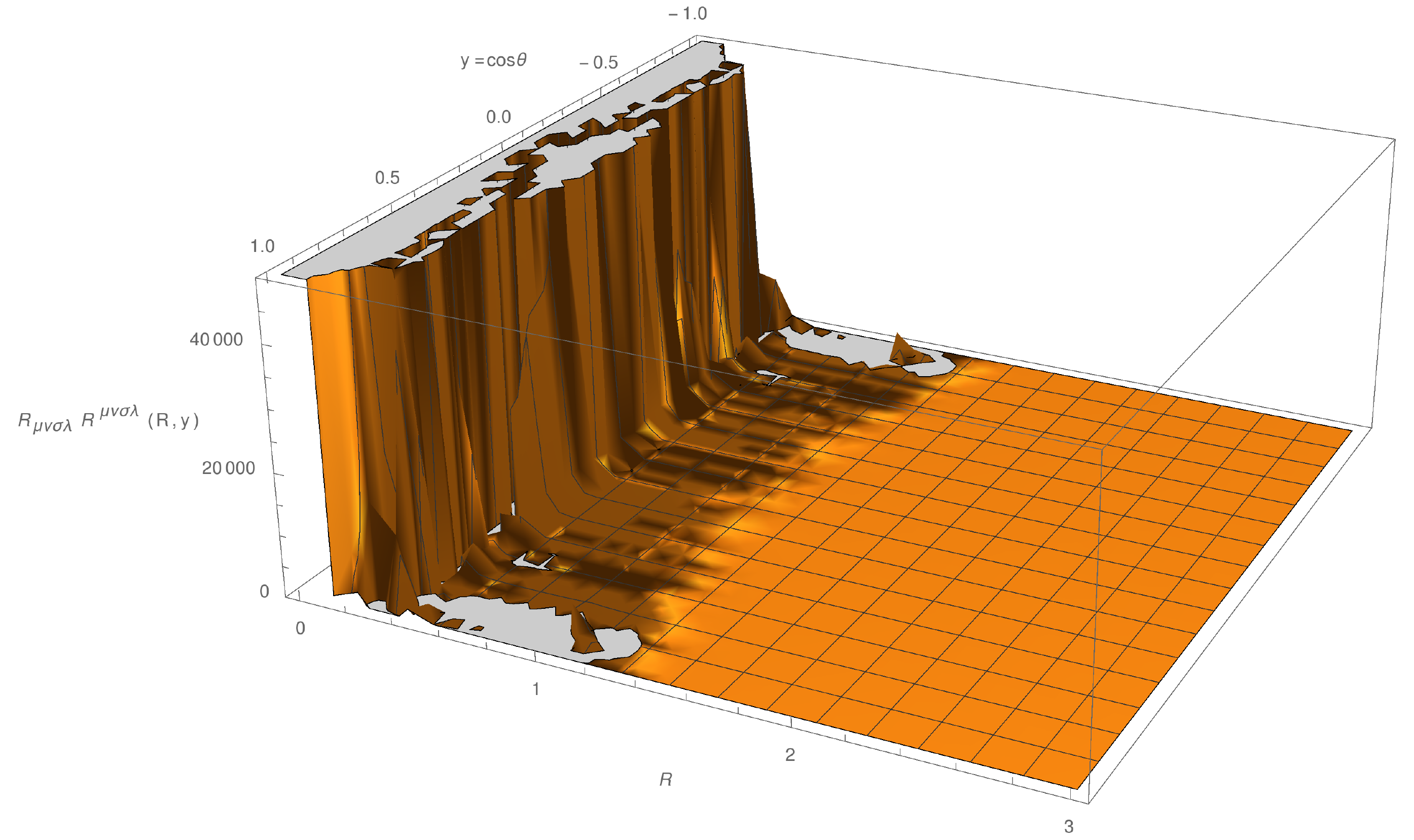} 
  \caption{Plot of the $\textrm{R}_{\m\n\s\l} \textrm{R}^{\m\n\s\l}(R,y) $ curvature invariant for particular values of the parameters $\d = 1/2 ,  \a = 1/2 , \k=1$. It diverges when the radial coordinate $R \rightarrow 0$. This behaviour remains qualitatively the same for other values of $\a , \k$  in the range $0\le \a \le 1$ and $\k \ge 0$.}
\label{figu1} 
 \end{figure}
 The surface horizon area, defined by $R=\k/2$,  is given by
 \beq
             S_H = \int_0^{2\pi} d \varphi \int_{-1}^1 dy \sqrt{g_{yy} g_{\varphi \varphi}} \ \  \Big|_{R=m} =\pi \k^2  \quad .
\eeq
Therefore, similarly to the standard GR case where the Kerr's event horizon area is given by $8\pi m(m+\sqrt{m^2-a^2})$, the presence of the rotation shrinks the size of the horizon, for a given value of the mass. Nevertheless its geometry remains spherical as in the static case, this can be understood by looking at the equatorial and polar circumferences, which respectively are
\bea
          C_e  &=&  \int_0^{2\pi} \sqrt{g_{\varphi\varphi}} d\varphi   \ =  \pi \k    \ \  ,  \\
           C_p  &=& \int_{-1}^1 \sqrt{g_{yy}} dy   \ =  \pi \k  \ \ .
\eea
The  topology of the $\mathcal {S}_h$ surface can be checked with the help of the Gauss-Bonnet theorem. The Euler characteristic is given by
\beq
           \chi ( \mathcal{S}_h) = \frac{1}{2 \pi} \int_{0}^{2\pi} d\varphi  \int_{-1}^1 dy \sqrt{\bar{g}} \ \bar{\textrm{R}} = 2  \qquad ,
\eeq
where $\bar{g}$ and $\bar{\textrm{R}}$ are the determinant and Ricci scalar curvature of the metric defined on the surface's horizon $\mathcal{S}_h$ at constant time. Therefore the genus $\mathfrak{g}=\chi(\mathcal{S}_h)/2-1$ of the surface $\mathcal{S}_h$ is null, so the horizon topology is spherical.\\
The radial coordinate (\ref{coor-transf}) was chosen as the simplest one containing the static radial coordinates (\ref{xR}), in the limit of null rotation. But a better-suited coordinate transformation $x(R)$ might exist for describing  the stationary spacetime, in particular for a black hole interpretation.\\ 
Thanks to the Yamakazi potentials \cite{yamazaki-81} it is possible to write the spacetime defined by the Ernst potential (\ref{ernst-potential})  in a  closed metric form with the parameters $\d, \a , \k$  free. This is useful to recognise directly the limits to some notable spacetimes such as Schwarzschild, Kerr or BBMB. Thus, when the scalar field is conformally coupled and for $\a=\b$ (consequently, according to (\ref{tau-yama}), $\tau=0$), the structure functions in the metric (\ref{lwp-conf}) become\footnote{A Mathematica notebook with this metric can be found at \url{https://sites.google.com/site/marcoastorino/papers/1412-3539}.}
\bea
     f  &=& \frac{R_+ }{L_+} \left( \frac{x-1}{x+1} \right)^{\d-1} \label{f-rot-bbmb}  \quad ,  \\
     \omega  &=&   \k_1 - 2 \k \frac{M_+}{R_+}  \left( \frac{x-1}{x+1} \right)^{1-\d}  \quad ,  \\
   \g &=&  \mezzo \log \left[  \frac{\k_2 \ R_+}{x^2-y^2}  \right] \label{g-kad} \quad , \\
   \Psi &=& \sqrt{\frac{6}{8\pi G}} \tanh\left[\sqrt{\frac{1-\d^2}{12}} \log\left(\frac{x-1}{x+1} \right) \right] \quad , \label{Psi-del}
\eea
where 
\bea
         R_+(x,y) &=& (x^2-1)(1-\l\m)^2 - (1-y^2)(\l+\m)^2   \quad , \qquad  \\
         L_+ (x,y) &=& (1-\l\m) \left[(x+1)^2-\l\m(x-1)^2\right] + (\l+\m) \big[\l(1-y)^2+\m(1+y)^2 \big]  \quad , \\
         M_+(x,y) &=& (x^2-1) (1-\l\m) \big[\l+\m-y(\l-\m)\big] + (1-y^2)(\l+\m) \big[1-\l\m+x(1+\l\m)\big] \  , \qquad \quad \label{M+}
\eea
while the conformal factor $\Omega$ is given by (\ref{conf-tr}), $\l(x,y)$ and $\m(x,y)$ are the same of (\ref{lambda}) - (\ref{mu}) respectively.
The integration constants $\k_1$ and $\k_2$ are fixed by requiring elementary asymptotic flatness of the metric (\ref{f-rot-bbmb})-(\ref{Psi-del}) as follows 
\bea
           \lim_{x\rightarrow\infty} \omega = 0 \quad &\Longrightarrow & \qquad \k_1=\frac{4 \  \k \ \a}{\a^2-1} \ \ \label{k1}, \\
           \lim_{y\rightarrow\pm1} \g= 0 \quad &\Longrightarrow & \qquad   \k_2=(\a^2-1)^{-2}   \ \ ,
\eea
while $\k$ remains the same of (\ref{yama-k}). The constraint (\ref{k1}) for $\k_1$ also arise demanding the regularity of the metric on the rotation axis. In fact according to \cite{carter}  $g_{\varphi\varphi}$ and $ g_{t\varphi}$ have to vanish where the killing vector $\p_\varphi=0$. The main difference with respect to standard general relativity \cite{quevedo},  appears in $\g(x,y)$ which in our  case, according to (\ref{gr})-(\ref{gipsi}), assumes the simple expression (\ref{g-kad}). Actually when $\d=1$ the scalar field vanishes so, for that value, we recover the rotating black hole of Einstein theory: the Kerr spacetime.\\
Some limits to notable spacetime are shown in the following table \ref{tabella}.

\begin{table}[h]  \label{tabella}
\begin{center}
\begin{tabular}{|c|c|c|c|c|} 
 \hline
Space-Times & $\a=\b$ & $\k$ & $\d$ \\ 
\hline 
\rule[-1ex]{0pt}{2.5ex} Kerr  Black Hole  & $  \pm \displaystyle \sqrt{\frac{m-\sqrt{m^2-a^2}}{m+\sqrt{m^2-a^2}}} $ & $ \sqrt{m^2-a^2} $ & 1 \\ 
\hline 
\rule[-1ex]{0pt}{2.5ex}  Schwarzschild  Black Hole & 0  & $m$ & 1 \\ 
\hline 
\rule[-1ex]{0pt}{2.5ex} BBMB Black Hole & 0 & $ 2 m $   & 1/2 \\
\hline 
\rule[-1ex]{0pt}{2.5ex} Rotating BBMB  & $\a$ & $ 2m \displaystyle \frac{(1-\a^2)}{1+3\a^2} $   & 1/2 \\
\hline 
\end{tabular}  
 \caption{Some specialisation of the metric (\ref {lwp-conf}),(\ref{f-rot-bbmb})-(\ref{M+}), for some values of its parameters. $m$ and $a$ denote the standard mass and angular momentum (for mass unit) of the Kerr spacetime.} \label{table1}
\end{center}
\end{table}

In order to have the Kerr spacetime in the standard Boyer-Lindquist coordinates representation just define
\beq
         x = \frac{r-m}{\k}    \qquad , \qquad  \qquad  y = \cos\theta \quad .
\eeq
While to recover the static BBMB black hole (\ref{bbmb-bh}) one has to use the coordinate transformation (\ref{xR}).\\
Even though the distortion parameter $\d$ continuously connects the Kerr black hole with the rotating version of the BBMB black hole we do not expect to have a physical process that actually connects these two black holes. That's because even in the static limit when $1/2 < \d < 1$ one has naked singularities.  \\
Note that these spacetime are naturally nut free (because $\a=\b$) but is possible to add NUT charge with an extra Ehlers transformation, as we have done to obtain the more general case (\ref{ersnt-pot-mono}).

\section{Multipolar FJRW metrics}
\label{multi-section}

It is possible to push 	further the solution generating mechanism with the minimally and conformally coupled scalar field to construct mass and angular multipolar generalisation of the FJRW solutions with an infinite number of independent parameters.   We recall that the mass multipole solutions have the peculiar property that they do not vanish in the Newtonian limit, unlike the angular multipoles (i.e. the ones carried by the Tomimatsu-Sato solution). On the other hand, the angular multipoles are produced by the mass deformation of the body due to the rotation. The simplest example, in case of null scalar field, is given by the Erez-Rosen metric which is a static spacetime endowed with a quadrupole moment. Of course these solutions in general have curvature singularities not covered by an event horizon, therefore are not suitable to describe black holes, but they can describe other astrophysical objects. By applying the HKX transformation it is  possible to build new exact stationary and axisymmetric vacuum solutions possessing an arbitrary large number of independent parameters \cite{quevedo}. \\
 These results can be directly generalised to the case of a minimally or conformally coupled scalar field as we have done in the monopolar solutions of sections \ref{b=0} and \ref{secB}. To do so one has to generalise (\ref{lambda})-(\ref{ersnt-pot-mono}) to:
\bea
          \bar{\l} &=&  \a (x^2-1)^{1-\d} (x+y)^{2\d-2} \exp\left[ 2\d \sum_{n=1}^\infty (-1)^n q_n B_{n-} \right]  \label{lambda-bar} \quad , \\   
          \bar{\mu} &=&  \b (x^2-1)^{1-\d} (x-y)^{2\d-2}  \exp\left[ 2\d \sum_{n=1}^\infty (-1)^n q_n B_{n+} \right] \quad ,  \\
          \bar{\Er}   &=& \frac{(d_++d_- e^{2\d\psi}) e^{i\tau} - (d_+-d_-e^{2\d\psi}) }{(d_++d_-e^{2\d\psi}) e^{i\tau}+(d_+-d_-e^{2\d\psi}) }  \label{Er-bar} \quad , 
\eea
where, for $n \geq 0$,
$$      B_{n\pm} = \frac{(\pm1)^n}{2} \log \left[ \frac{(x\mp y)^2}{x^2-1} \right]  - (\pm 1)^n Q_1(x) +  P_n(y) Q_{n-1}(x) - \sum_{k=1}^{n-1} (\pm1)^k P_{n-k}(y) [ Q_{n-k+1}(x)-Q_{n-k-1}(x) ]  \ \ ,  $$
\beq \label{spi-gen}
          \psi(x,y) =  \sum_{n=1}^\infty (-1)^{n+1} q_n P_n(y) Q_n(x) 
\eeq
where $ P_n(y) $ are the Legendre polynomials and $ Q_n(x)$ are the Legendre functions of the second kind\footnote{See appendix \ref{app-legendre} for more information about the Legendre functions of the second kind.}; $d_\pm$ follows the definition (\ref{dpm}). $\{q_n\}_{n=0,1,2,...}$ are independent constants related to the metric multipolar expansion, for both angular or mass multipole moments. To be more precise, the $q_n$ term  gives contributions to the $2^n$ multipole, further details can be found in appendix (\ref{app-mome}) or in \cite{quevedo}. Here integration constants are set to zero according to
$$ \lim_{x\rightarrow\infty} B_{n\pm} = 0 \quad . $$
In sections \ref{b=0} and \ref{secB} we have considered the simplest case where $q_0=1$ and $q_j=0 \ \forall \ j >0 $, in that case equations (\ref{lambda-bar})-(\ref{Er-bar}) trivially reduced to  (\ref{lambda})-(\ref{ersnt-pot-mono}). \\
Up to this point the Ernst potential has worked well for both the vacuum case, describing stationary rotating multipolar Zipoy-Woorhees metrics, or, for the scalar coupling, describing stationary rotating FJRW metrics. From the Ernst potential we can extract the $f(x,y)$ and $\om(x,y)$ fields. But the main difference in the two theories consists in the remaining $\g(x,y)$ structure function of the Lewis-Weyl-Papapetrou metric, and a further possible conformal transformation if we want to work in the conformally coupled theory. To obtain $\g(x,y)$ one has to integrate the equations (\ref{gr})-(\ref{gz}), where the presence of a non-trivial scalar field becomes relevant. For the scalar field (\ref{scalar-mini}) mainly considered in this paper the correction with respect to standard general relativity is given in  (\ref{gipsi}).\\
As a significant example we will now build the Erez-Rosen metric with a minimally coupled scalar field. The standard Erez-Rosen metric can be built from equations (\ref{lambda-bar})-(\ref{spi-gen}) fixing the parameters as follows:
\beq
      q_0 = 1 \quad , \quad q_1=0 \quad ,\quad q_2 \neq 0 \quad , \quad q_j = 0 \ \ (j>2) \qquad , \qquad \k=m  \quad , \qquad \a = \b = \tau= 0  \quad , \qquad \d=1 \nn 
\eeq
Analogously if we want to have a Erez-Rosen metric in presence of a minimally (or conformally) coupled scalar field (\ref{scalar-mini}) (or (\ref{scalar-conf})) we have to choose the same values for the parameters of the vacuum case $q_j,\k,\a,\b$, so that  asymptotically and in the weak field limit, for small $m$, the scalar coupled cases have a similar multipolar behaviour with respect to the vacuum case. Obviously in this case $\d=1/2$ because the metric has to reduce to FJRW (or BBMB) spacetime when the quadrupole moment of the source vanishes (i.e. $q_2=0$), in the same way the Erez-Rosen metric reduces to the Schwarzschild black hole. With this parametric imposition the Ernst potential (\ref{Er-bar}) becomes
\beq
          \Er = f = \exp \left\lbrace q_2 (3 y^2 -1) \left[ \frac{1}{4} \left( 3x^2-1 \right) \log\left( \frac{x-1}{x+1}  \right) + \frac{3}{2} x \right] \right\rbrace \sqrt{\frac{x-1}{x+1}} \quad .
\eeq
Since the spacetime is static, the Ernst potential is not complex and $\omega=0$, therefore the remaining unknown function can be obtained by integrating  (\ref{gr}) and (\ref{gz}), to get
\bea
      \g &=&
\frac{1}{2} \left(1+\frac{q_2}{2}+\frac{q_2^2}{4}\right) \log \left(\frac{x^2-1}{x^2-y^2}\right) + \frac{9q_2^2}{256} \left(x^2-1\right) \left(y^2-1\right) \left[x^2 \left(9 y^2-1\right)-y^2+1\right] \log ^2\left(\frac{x-1}{x+1}\right) \nn \\
&+& \frac{3q_2}{64}\left(y^2-1\right) \left\{\left[1+x \log \left(\frac{x-1}{x+1}\right)\right] \Big[8+q_2\left[3 \left(9 x^2-7\right) y^2-3 x^2+5\right]\Big]+\Big[8+q_2 \left(9 y^2-1\right)\Big]\right\} \ \ .\nn
\eea
Here the arbitrary integration constant was set to fulfil (\ref{lit-g}) to avoid conical singularities on the symmetry axis. The scalar field remains as in (\ref{scalar-mini}) or (\ref{scalar-conf}) depending if we are considering the Einstein or Jordan frame respectively.
Let's compute the first mass and angular multipoles moments for the above specetime. Using the general results of appendix \ref{app-mome} we have, for the minimally coupled system
\bea
        M_0 &=&  m \qquad , \qquad M_1 = 0 \qquad , \qquad  M_2 = m^3 \left( 1 + \frac{8}{15} q_2 \right)  \qquad ,  \\ 
        J_j &=&  0 \qquad \ , \qquad  \forall \  j \geq 0  \quad .
\eea
There is a difference with respect to the Erez-Rosen mass multipole moments, basically due to the different value of the Zipoy parameter $\d$, as for instance it can be seen by looking at the mass quadrupole moment (the Erez-Rosen value is $M_2^{ER}=2q_2m^3/15$).\\

\section{Comments and Conclusions}

In this paper the Ernst solution generating technique, in the context of standard Einstein gravity with a (minimally or) conformally coupled scalar field, is enhanced to include the HKX transformations. These transformations are able to add rotation meanwhile preserving  asymptotic  and elementary flatness.  Applying these methods we were able to generate a large family of asymptotically flat, axisymmetric and stationary solutions for both the minimally and the conformally coupled theory, containing, apart the Zipoy-Woorhees-distortion parameter $\d$ and the mass $m$,  two independent parameters, the rotation and reflection parameters $\a$ and $\b$. We explain how to remove the possible NUT charge emerging from the HKX transformation. As significant examples we analysed some special cases, that are continuously connected to the Kerr black hole by the distortion parameter, where only one independent extra parameter was left: the rotation (i.e. $\b=0$ and $\a=\b$). In the minimal frame they can be considered as the stationary extension of the Janis, Winnicour, Robinson and Fisher solution, while in the conformally coupled theory they include a rotating generalisation of the BBMB black hole. 	Although both cases have a clear limit to the BBMB black hole when turning off the rotation parameter, the case $\a=\b$ is the most similar to the rotating black hole in GR, that is,  an angular and mass multipolar expansion and geometry similar to the extremal Kerr spacetime.  Depending on the relative values of the $\a$ and $\b$  parameters, introduced by the HKX transformation, these axisymmetric spacetimes can be symmetric with respect to the equatorial plane or not. The more general case where both the rotation and reflection parameters are not null and independent remains to be studied.\\
This family has been further generalised to contain an arbitrary number of independent parameters related to additional mass multipoles. As an example we provide an Erez-Rosen like spacetime in the presence of a scalar field.   \\
Note that the static seed metric of the BBMB black hole coincides with that of the extremal Reissner-Nordstrom black hole. Therefore if one wants to apply the Janis-Newman (JN) algorithm for adding rotation,  the extremal Kerr-Newman metric would be  obtained, which is not a solution for the theory we are dealing with. This occurs because the JN algorithm was discovered, a posteriori, to work within Einstein-Maxwell general relativity and it is just a (complex) coordinate transformation, thus not dependent on the specific theory one is actually considering. On the other hand the resulting stationary metrics we have built, after the HKX transformation in the Ernst formalism, are different from the Kerr-Newman, and they are proper solutions of the field equations. \\ 
It may also be interesting, for a future perspective,  to add the cosmological constant term, because it turned out to be  useful in regularising the behaviour of the scalar field on the horizon. That's because the cosmological constant (of the appropriate positivity) shifts the position of the horizon so that the divergence of the scalar field is protected by the event horizon \cite{martinez}. Of course this is not a trivial task since a solution generating technique that includes the cosmological term is not known at the moment \cite{marcoa-lambda}. \\
HKX transformations can be adapted in other gravity theories connected to general relativity with a minimally coupled scalar field by a conformal transformation, such as Brans-Dicke or some $f(R)$ gravity, basically in the same way as described in this paper for general relativity with a conformally coupled scalar field.
\\

%Note that the static seed metric of BBMB black hole coincides with the one of the extremal Reissner-Nordstrom black hole. Therefore if one wants to apply a not theory dependent algorithm, such as the Janis-Newman (JN) one, for adding rotation, he would obtain the extremal Kerr-Newman metric, which does not solve the field equations for the theory we are dealing with. This because that JN algorithm was discovered, a posteriori, to work within Einstein-Maxwell general relativity and is just a (complex) coordinate transformation, thus not dependent on the specific theory one is actually considering. On the other hand the resulting stationary metrics we have built, after the HKX transformation in the Ernst formalism, are different from the Kerr-Newman, and are proper solutions of the field equations. \\ 

\section*{Acknowledgements}
\small I would like to thank Eloy Ayon-Beato, Fiorenza de Micheli, Mokhtar Hassaine, Cristian Erices,  Hideki Maeda and Cristi\'{a}n Mart\'{i}nez for fruitful discussions. 
\small This work has been funded by the Fondecyt grant 3120236. The Centro de Estudios Cient\'{\i}ficos (CECs) is funded by the Chilean Government through the Centers of Excellence Base Financing Program of Conicyt.
\normalsize

\appendix

\section{Legendre polynomials and functions of the second kind}
\label{app-legendre}

Legendre polynomials $P_n(x)$ can be obtained by the Rodrigues formula
\beq
       P_n(x) = \frac{1}{2^n n!} \ \frac{d^n}{dx^n} [ (x^2-1)^n ] \quad .
\eeq
we list the firsts
\bea
       P_0(x) &=& 1     \quad ,                    \\
       P_1(x) &=& x        \quad ,                            \\
       P_2(x) &=& \mezzo (3x^2-1)    \quad ,             \\
       P_3(x) &=&   \mezzo (5x^3 - 3x)        \quad ,           \\
      P_4(x) &=&  \frac{1}{8} (35x^4 -30 x^2 + 3) \quad .
\eea
Legendre functions of the second kind $Q_n(x)$ can be built by means of $P_n(x)$  with the following prescription
\beq
        Q_n(x) = \mezzo P_n(x) \log\left( \frac{x+1}{x-1} \right) - W_{n-1}(x)  \quad ,
\eeq 
where
\beq 
       W_{n+1} = \sum_{k=1}^n \frac{1}{k} P_{k-1} (x) \ P_{n-1}(x) \quad ,
\eeq
thus the firsts are
\bea
       Q_0(x) &=& \mezzo \log\left( \frac{x+1}{x-1} \right)                \quad ,     \\
       Q_1(x) &=& \mezzo x \log\left( \frac{x+1}{x-1} \right) -1          \quad ,       \\
       Q_2(x) &=& \frac{1}{4} (3x^2-1) \log\left( \frac{x+1}{x-1} \right) - \frac{3}{2} x                             \quad ,                                                \\
       Q_3(x) &=&   \frac{1}{4} (5x^3-3x) \log\left( \frac{x+1}{x-1} \right) - \frac{5}{2} x^2 + \frac{2}{3}          \quad .        
\eea

\section{Cosgrove's metrics with a scalar field}
\label{secC}

For sake of completeness we also present the extension of another solution generating technique, based on Ernst equations and complex potentials, able to achieve stationarity without spoiling the asymptotic flatness, given by Cosgrove in \cite{Cosgrove-new1} and \cite{Cosgrove-new2}. It provides the rotating generalisation of the Zipoy-Woorhess metric and the generalisation of the Tomimatzu-Sato for not integer parameter $\d$ inequivalent with respect to the sections \ref{b=0}, \ref{secB} and \ref{multi-section} which are based on the HKX transformation. It is enough concise to work directly, for a generic $\d$, in the metric formalism, not only in the Ernst picture. Let's begin considering an example containing both the Kerr and the Zipoy-Woorhess metrics. We will present the standard separable Cosgrove solution of \cite{Cosgrove-new2} and we will show how to adapt it to the presence of the scalar field according to (\ref{gr-psi})-(\ref{gz-psi}). It can be most compactly expressed when the NUT charge is not null, further on we will show how to remove it, whether desired. When the scalar field is null the axisymmetric stationary metric is given by the following Ernst potential
\beq \label{cos-pot}
     \Er= \frac{d_+-d_-}{d_++d_-} 
\eeq
with
\beq
      d_\pm = \frac{p}{2} (x^2-1)^{\bar{\d}} \Big[ (x+1)^{\bar{\d}+1} (1-y)^{\bar{\d}} \pm (x-1)^{\bar{\d}+1} (1+y)^{\bar{\d}} \Big] + \frac{iq}{2} (1-y^2)^{\bar{\d}} \Big[ (x+1)^{\bar{\d}} (1-y)^{\bar{\d}+1} \mp (x-1)^{\bar{\d}} (1+y)^{\bar{\d}+1} \Big] \ ,
\eeq
where $p$ and $q$ are two dependent parameters related to the mass and the angular momentum:  when $q=0$ the Ernst potential remains real, so the metric is static; they are related by the usual constraint $p^2+q^2=1$. $\bar{\d}$ is chosen to fit the notation of \cite{Cosgrove-new2} and is related with ours by $\d=\bar{\d}+1$, hence the Kerr spacetime is now given  for $\bar{\d}=0$. Note that for $-1 \leq \bar{\d} \leq 0$  (or $ 0 \leq \d \leq 1$) the scalar field is real, while otherwise imaginary.  \\
Explicitly the Ernst potential (\ref{cos-pot}) has the form 
\beq
        \Er =  \left[\frac{(x-1)(1+y)}{(x+1)(1-y)} \right]^{\bar{\d}} \frac{p(x^2-1)^{\bar{\d}}(x-1) -iq(1-y^2)^{\bar{\d}}(1+y) }{p(x^2-1)^{\bar{\d}}(x+1) -iq(1-y^2)^{\bar{\d}}(y-1) } \quad .
\eeq
Note that  this potential does not contain the static BBMB spacetime, therefore can not considered a good seed neither for a stationary BBMB.\\
The structure functions of the Lewis-Weyl-Papapetrou metric descending from the potential (\ref{cos-pot}) are
\bea
        f &=&  \left[\frac{(x-1)(1+y)}{(x+1)(1-y)} \right]^{\bar{\d}} \frac{p^2(x^2-1)^{2\bar{\d}+1} -q^2(1-y^2)^{2\bar{\d}+1}}{p^2(1+x)^2(x^2-1)^{2\bar{\d}} +q^2(1-y)^2(1-y^2)^{2\bar{\d}} } \label{f-cos} \quad , \\
      %  h &=&  - \frac{2pq\big[ (x-1)(1+y) \big]^{2\bar{\d}} (x+y) }{p^2(1+x)^2(x^2-1)^{2\bar{\d}} +q^2(1-y)^2(1-y^2)^{2\bar{\d}} } \\
     \om   &=&  -\k \frac{2pq\big[ (x+1)(1-y) \big]^{2\bar{\d}+1} (x+y) }{p^2(x^2-1)^{2\bar{\d}+1} -q^2(1-y^2)^{2\bar{\d}+1}}  \label{om-cos}  \quad ,\\
      e^{2\g_0}  &=& b \frac{(x^2-1)^{\bar{\d}^2} (1-y^2)^{\bar{\d}^2}}{(x-y)^{(2\bar{\d}+1)^2}(x+y)} \left[ p^2(x^2-1)^{2\bar{\d}+1} -q^2(1-y^2)^{2\bar{\d}+1} \right] \quad , \label{g-0}
\eea
where $b$ is an arbitrary integration constant. When the scalar field (\ref{scalar-mini}) is present the only structure function of the Lewis-Weyl-Papapetrou metric that changes is $\g$. It can be found thanks to (\ref{gr})-(\ref{gz}):
\beq \label{g-psi}
         e^{2\g_\Psi} = b \frac{ (x+y)^{\bar{\d}^2+2\bar{\d}-1} (1-y^2)^{\bar{\d}^2}}{(x-y)^{3\bar{\d}^2+2\bar{\d}+1} (x^2-1)^{2\bar{\d}} }\left[ p^2(x^2-1)^{2\bar{\d}+1} -q^2(1-y^2)^{2\bar{\d}+1} \right] = e^{2\g_{0}} \left( \frac{x^2-y^2}{x^2-1} \right)^{\bar{\d}^2+2\bar{\d}}
\eeq
For $\bar{\d}=0$ the scalar field is null, $ \g_\Psi \rightarrow \g $ and the spacetime becomes the Kerr-NUT black hole. We can remove the NUT charge by applying an Ehlers transformation to the Ernst potential of \cite{Cosgrove-new1} and requiring the appropriate falloff boundary conditions.  So we add an extra NUT charge, parametrised by $\tau$, as done in section \ref{sec-bbmb}, the Ehlers transformed Ernst potential (\ref{cos-pot}) is 
\beq
              \Er= \frac{d_+ e^{i\tau} -d_-}{d_+e^{i\tau}+d_-}  \quad .
\eeq
When $\bar{\d}=0$ the $\omega$ function coming from this potential is given by
\beq
           \omega= \omega_0 +\frac{2\k}{q} (p \cos \tau+q \sin \tau) - \frac{2 \kappa  p \left(x^2-1\right) \left[\left(p^2 x-q^2 y\right) \cos \tau  +p^2+ p \  q  (x+y) \sin \tau +q^2\right]}{p^2 q \left(x^2-1\right)+q^3 \left(y^2-1\right)} \  \ ,
\eeq
whose asymptotic behaviour for large $x$ is given by
\beq
\omega \approx \left(-\frac{2 \kappa  q}{ p}-\frac{2 \kappa  p}{ q} + \frac{2 \kappa   q y \cos \tau}{ p}-2 \kappa  y \sin \tau + \omega_0 \right) + \frac{2 \kappa  q \left(y^2-1\right) [ p \cos \tau + q \sin \tau ]}{ p^2 x} + O\left( \frac{1}{x^2} \right) 
\eeq
Requiring the usual falloff at spatial infinity $O(1/x)$ we impose 
\beq  \label{const-cos-nut}
         \cos  \tau = p  \qquad    \qquad    \text{and}    \qquad   \qquad    \om_0=\frac{2\k}{pq} \quad .
\eeq
Note that (\ref{const-cos-nut}) with $p^2+q^2=1$ implies that $\sin \tau =q$. With fine tuning of the NUT charge we have erased the previous existing one. Therefore we remain with a pure Kerr spacetime. To convince oneself of this it is sufficient to check the constrained Ernst potential which is exactly that of Kerr spacetime:
\beq
             \Er\  \Big|_{\bar{\d}=0}  \ = \ 1- \frac{2\ (p+iq)}{p+iq+e^{i\tau} (px-iqy) } \  \longrightarrow \  \frac{px-iqy-1}{px-iqy+1}  \qquad .
\eeq
%It is possible to find for each fixed $\bar{\d}$ the asymptotically flat version of  (\ref{f-cos}) and (\ref{om-cos}) with or without the scalar field in a compact way, but the general expression (depending explicitly from $\bar{\d}$) is pretty involved so we skip to write it. 
%Maybe we can write the asymptotically flat $f$ and $\om$, which are simplified by the constrains on $\tau$...
\\
For $\d>0$ the spacetimes (\ref{f-cos}) - (\ref{g-psi}) are NUT free, so we don't need an additional Ehlers transformation (but $\Psi$ becomes imaginary). \  On the other hand $\g$ and $\g_\Psi$ remain the same as before: (\ref{g-0}) and (\ref{g-psi}) respectively, because the Ehlers transformations do not affect eqs. (\ref{gr}) and (\ref{gz}) \cite{treves}.   \\

\section{Multipolar moments}
\label{app-mome}

It is possible to compute the angular and mass multipole moments, from the Ernst potential in prolate spheroidal coordinate \cite{fodor}, \cite{quevedo}. This can clarify the role of the independent constants that appear in the general multipolar metric presented in section \ref{multi-section}. There are several definitions of multipole moments for axisymmetric fields, we are considering here those of Geroch-Hansen \cite{hansen}. These have the advantages of being coordinate independent and they coincide with the Newtonian moments (in case of flat spacetime). \\
According to the notation used in (\ref{lambda-bar})-(\ref{Er-bar}) we will list the first mass $M_j$ and angular $J_j$ multipole moments (for more details see \cite{quevedo}\footnote{After the completion of this paper \cite{pappas} was published  where  the contribution of the scalar field is also taken into account.})%\footnote{Differences in the general mass quadrupole moment and mass dipole of Kerr-Penrose-Rosen metric with respect to \cite{quevedo} are just due to typos in \cite{quevedo}.}. 
\bea
       M_0 &=& \k \left( \d q_0 +\frac{2 \a \b}{1-\a\b} \right)\label{Mi} \\
       M_1 &=& \k^2 \left[ -\frac{\d q_1}{3} + \frac{\b^2-\a^2}{(1-\a\b)^2} \right]  \nn \\
       M_2 &=& \k^3 \left\lbrace \frac{2 \d q_2}{5}  - \frac{\d^3}{3} - \frac{2 \d^2 \a \b}{1-\a\b}  + \d \left[ \frac{1}{3} + \frac{\a\b(-2-2\a\b+3\a^2+3\b^2+4\a^2\b^2) - 3(\a^2+\b^2)}{(1-\a\b)^3}  \right]  \right.  \nn \\ 
  & & +  \left. 2 \  \frac{(\a+\b)^2-\a \b (1+2\a^2+2\b^2+2\a\b+\a^2\b^2) }{(1-\a\b)^3}      \right\rbrace \nn \\
     M_3 &=& -\frac{k^4 (\a^2-\b^2) \left\{\a^2 \left[3 \b^2 (\delta -1)^2-1\right]-2 \a \b [3 (\delta -2) \delta +4]-\b^2+3 (\delta -1)^2\right\}}{(\a \b-1)^4}\nn \\
     J_0 &=& -\k \frac{\a-\b}{1-\a\b} \label{Ji}\\
     J_1 &=& -\k^2 \frac{\a+\b}{(1-\a\b)^2} \left[ 3\a\b + 2\d (1-\a\b) -1 \right] \nn \\
     J_2 &=& -\frac{\k^3}{1-\a\b} \left[ - \frac{2}{3} \d q_1 (\a+\b) + (\a-\b)(1-\d)^2 \right] - \frac{\k^3}{(1-\a\b)^3} \left( \b^3- \a^3 + \a \b^2 - \a^2 \b \right) \nn \\
     J_3 &=&  \frac{\k^4 (\a + \b) \left[5 \a^3 \left(\b^3 + \b \right) + \a^2 \left(\b^2 - 3 \right) + \a \b \left(5 \b^2-3\right)-3 \b^2 + 1\right]}{(1-\a \b)^4}  \nn \\
       &+& \frac{\delta  \k^4 (\a+\b) \left\{\a^2 \left[\b^2 \left(2 \delta ^2-15 \delta +28\right)+12\right]-4 \a \b \left(\delta ^2-3 \delta -1\right)+12 \b^2+2 \delta ^2+3 \delta -8\right\}}{3 (1- \a \b)^3}  \nn
\eea
In $M_3$ and $J_3$ we put for simplicity $q_i=0 \ \forall i$.
When the NUT parameter $\tau \neq 0$ the angular and mass multipole moments, for $n\leq 3$, are modified as follows
\bea
       M'_n &=& M_n \cos \tau - J_n \sin \tau  \nn \\
       J'_n &=& M_n \sin \tau + J_n  \cos \tau \label{Mi-Ji-rot}
\eea 
The presence of odd mass multipoles and even angular multipole moments means that the metric is not symmetric with respect to the equatorial plane, $y=0$. Using (\ref{Mi})-(\ref{Mi-Ji-rot}) it is easy to obtain the first multipole moments for the Kerr black hole of sections \ref{b=0} and \ref{secB}
\bea
        M_0 &=&  m \qquad \ \ , \qquad M_1 = 0 \qquad \quad , \qquad \  M_2 = -m a^2  \qquad , \ \ \qquad M_3 = 0 \qquad ,\\ 
        J_0 &=&  0 \qquad \ \ \ , \  \qquad J_1 = am \qquad \ , \ \ \qquad  J_2 = 0 \qquad  \qquad  , \quad \qquad J_3 = - m a^3 \quad .
\eea
The monopole term is the mass of the black hole, while the angular dipole moment coincides with the angular momentum. The higher multipoles are due to the rotation and reflect the fact that the stationary Kerr black hole looses the spherical symmetry typical of the static Schwarzschild one.  \\

\section{More general scalar fields}
\label{app-app-expan-g}

The most general form for the scalar field in the minimal frame, that can be obtained by variable separation, is given by (\ref{scalar-general}).        
\beq
\hat{\Psi} = \sum_{n=0}^\infty \left[ a_n Q_n(x) + b_n P_n(x) \right] \left[c_n Q_n(y) + d_n P_n(y)  \right] 
\eeq
Applying the condition of asymptotic flatness we set to zero the coefficients  $b_n$ and $c_n$, and considering $\d=1/2$, the scalar field becomes 
\bea
            \hat{\Psi} &=& \sum_{n=0}^\infty a_n Q_n(x)  P_n(y)  \\
                            &=&  \frac{a_0}{2} \log \left( \frac{x-1}{x+1} \right) + a_1 \left[ \frac{x}{2} \log\left(  \frac{x-1}{x+1} \right) +1 \right] y + a_2 \left[ \frac{3x^2-1}{4}\log\left(  \frac{x-1}{x+1} \right) +  \frac{3}{2} x \right] \left( \frac{3y^2-1}{2} \right) + ... \nn
\eea  
We can evaluate the contribution of the scalar's first terms expansion to the $\g=\g_0+\sum_{n=0}^\infty \g_{\Psi_n}$ field. According to (\ref{gr-psi})-(\ref{gz-psi}) the first contributions  are given by
\bea
         \g_{\Psi_0} &=& c_0 + \frac{a_0^2}{4} 8\pi G  \log\left( \frac{x^2-1}{x^2-y^2} \right) \quad , \\
          \g_{\Psi_1} &=& \frac{a_1^2}{16} 8\pi G \left\lbrace 4\log\left( \frac{x^2-1}{x^2-y^2} \right) + (y^2-1) \log \left( \frac{x-1}{x+1} \right)  \left[ 4x + (x^2-1) \log\left( \frac{x-1}{x+1} \right)  \right] \right\rbrace  \quad , \\
           \g_{\Psi_2} &=& \frac{a_2^2}{32}8\pi G \left\lbrace 8 \log\left( \frac{x^2-1}{x^2-y^2} \right) +9x^2 +6y^2(8-15x^2) + 9y^4(9x^2-4) +\frac{3}{4}(y^2-1)\log\left( \frac{x-1}{x+1} \right)  \cdot \right. \nn \\
           & & \left. \cdot \left[4x[5-3x^2+3(9x^2-7)y^2]+ 3(x^2-1)[1-x^2+(9x^2-1) y^2]\log\left( \frac{x-1}{x+1} \right)  \right] \right\rbrace  \ \ , 
\eea
where $c_0$ is an integration constant that can be fixed by physical requirements, such as, for instance, the absence of conical singularities.\\
If we we relax a little the boundary conditions allowing a constant falloff of the scalar field also the coefficient $b_0$ can be turned on. The effect of a non-null $b_0$ represent just a constant shift of the scalar field in the minimally coupled theory, which is a symmetry in the action (\ref{minimal-action}), but it reflects non-trivially in the conformally coupled theory. In fact starting from any seed solution  ($ds^2_0, \Psi_0$) of the conformally coupled theory it  is possible to obtain a  nonequivalent new solution in this way 
\bea
        \label{tras-ds}          ds^2_0  & \longmapsto & ds^2 =  \displaystyle \frac{1-\frac{8\pi G}{6} \Psi_0^2}{1-\frac{8\pi G}{6} \Psi^2} \ ds^2_0 \\
        \label{tras-psi}           \Psi_0       & \longmapsto &   \Psi =   \sqrt{\frac{6}{8 \pi G}}  \tanh \left\{ \sqrt{\frac{8 \pi G}{6}} \left[ b_0 + \sqrt{\frac{6}{8\pi G}} \ \text{arctanh} \left(\sqrt{\frac{8\pi G}{6}} \Psi_0 \right) \right]  \right\}            
\eea
These transformations, parametrised by the real number $b_0$, map solutions of the theory of General Relativity with a conformally coupled scalar field onto itself. In particular  when the seed metric is the BBMB black hole  (\ref{bbmb-bh}) we obtain after the transformation (\ref{tras-ds})-(\ref{tras-psi})
\bea
      ds^2 &=& \label{conf-metr} \frac{\big[\r(s+1)-2ms \big]^2}{4s \big[\r-m \big]^2} \left[  -\left( 1-\frac{m}{\r} \right)^2 dt^2   +  \frac{d \r^2}{ \displaystyle \left( 1-\frac{m}{\r} \right)^2}  + \r^2 d\theta^2 + \r^2  \sin^2\theta \ d\varphi^2  \right] \quad , \\
    \label{conf-psi}  \Psi &=& -\sqrt{\frac{8\pi G}{6}} \ \frac{\r(s-1)-2ms}{\r(s+1)-2ms} \quad ,
\eea
where for simplicity we have defined the parameter $ b_0 =  \sqrt{\frac{8\pi G}{6}} \mezzo \log s $. Of course when the parameter $b_0$ vanishes (so $s=1$) the transformation  (\ref{tras-ds})-(\ref{tras-psi}) becomes the identity and we recover the standard BBMB black hole (\ref{bbmb-bh}). On the other hand for non-null $b_0$ the transformation is not trivial as can be seen, for instance, looking at the contribution of the $s$ parameter in the scalar curvature invariants.\\
This solution was first found, by direct integration, in \cite{visser} and interpreted as a traversable wormhole. In the case where the cosmological constant is not null, the constant shift in the scalar field have the effect to map, in the action, the conformal scalar potential from a quartic power\footnote{Generically an additional scalar potential proportional to $\Psi^4$ can be considered in the action (\ref{action}) without spoiling the conformal invariance. It is not compatible with HKX transformations and usually it becomes relevant in presence of the cosmological constant. For these reasons it is not taken into account in the present work.} to a quartic polynomial, for further details see \cite{Ayon-Beato:2015b}.  Recently a solution to this system was found in \cite{anabalon-cisterna}. It admits a black hole interpretation and   generalises \cite{visser} to the presence of the cosmological constant. \\

\end{document}